\documentclass{article}
\usepackage{diagbox} 
\usepackage{float}
\usepackage{PRIMEarxiv}
\usepackage{xcolor}
\usepackage[utf8]{inputenc} 
\usepackage[T1]{fontenc}    
\usepackage{hyperref}       
\usepackage{url}            
\usepackage{booktabs}       
\usepackage{amsfonts}       
\usepackage{nicefrac}       
\usepackage{microtype}      
\usepackage{lipsum}
\usepackage{fancyhdr}       
\usepackage{graphicx}       
\graphicspath{{media/}}     

\title{Simulating the dynamics of NV$^-$ formation in diamond in the presence of carbon self-interstitials.}

\author{
  Guangzhao Chen, Joseph C.A. Prentice and Jason M. Smith* \\~\\
  Department of Materials, University of Oxford, Oxford, OX1 3PH, United Kingdom \\
  \\~\\
 \textit{guangzhao.chen@wolfson.ox.ac.uk}\\
 \textit{jason.smith@materials.ox.ac.uk} \\
}

\begin{document}
\maketitle

\begin{abstract}
This study utilises linear-scaling density functional theory (DFT) and develops a new machine-learning potential for carbon and nitrogen (GAP-CN), based on the carbon potential (GAP20), to investigate the interaction between carbon self-interstitials and nitrogen-vacancy (NV) centres in diamond, focusing on their excited states and diffusion behaviour. 
From the simulated excited states, 'Bright', 'Spike', and 'Dark' defect configurations are classified based on their absorption spectrum features. Furthermore, machine learning molecular dynamics simulation provides insight into the possible diffusion mechanism of C$_i$ and NV, showing that C$_i$ can diffuse away or recombine with NV. The study yields new insight into the formation of NV defects in diamond for quantum technology applications.

\end{abstract}


\section{Introduction}
The engineering of point defects in crystals at the single defect level is becoming increasingly important for the development of quantum technologies. For example, the negatively charged nitrogen-vacancy (NV$^-$) defect in diamond provides an optically addressable spin with long coherence times, even at room temperature, and has therefore become an important tool for quantum sensing \cite{degen2008scanning} and nanoscale imaging \cite{balasubramanian2008nanoscale} and is a promising candidate for quantum computing and quantum communication \cite{childress2013diamond, ovartchaiyapong2014dynamic,abobeih2022Nature}. Various other colour centre defects are also being explored, providing a rich source of new science and technological potential. Within this context, understanding and refining the processes by which these defects are created is of utmost importance.

Various methods are utilised for the controlled fabrication of colour centres, including ion implantation \cite{haque2017overview}, electron-beam irradiation \cite{schwartz2012effects} and laser writing \cite{chen2017laser}, with the general objective of introducing the necessary impurities and vacancies and facilitating their binding together within the crystal lattice to form the desired defect \cite{smith2019nanophotonics}. The creation of a vacancy in a crystal lattice is usually achieved by dislodging an atom from its lattice site into an interstitial location, thus creating a Frenkel defect - a vacancy-interstitial pair. The role of self-interstitials in the fabrication process, and their effect on the properties of the colour centres formed, therefore needs to be understood.

Here we focus on the dynamics of the formation of NV centres and model the effect of a migrating carbon self-interstitial (C$_i$) on the properties of a pre-formed NV centre. We compare our theoretical results with experimental data from the laser-writing of single NV centres, which has proven to be an effective technique both for creating Frenkel defects in diamond with minimal residual damage \cite{chen2017laser} and for the subsequent annealing step that leads to the formation of the NV complex \cite{chen2019laser}, with real-time monitoring of the NV fluorescence during the laser annealing process providing a window to the local dynamics.

\begin{figure}[H]
  \centering  \includegraphics[scale=0.6]{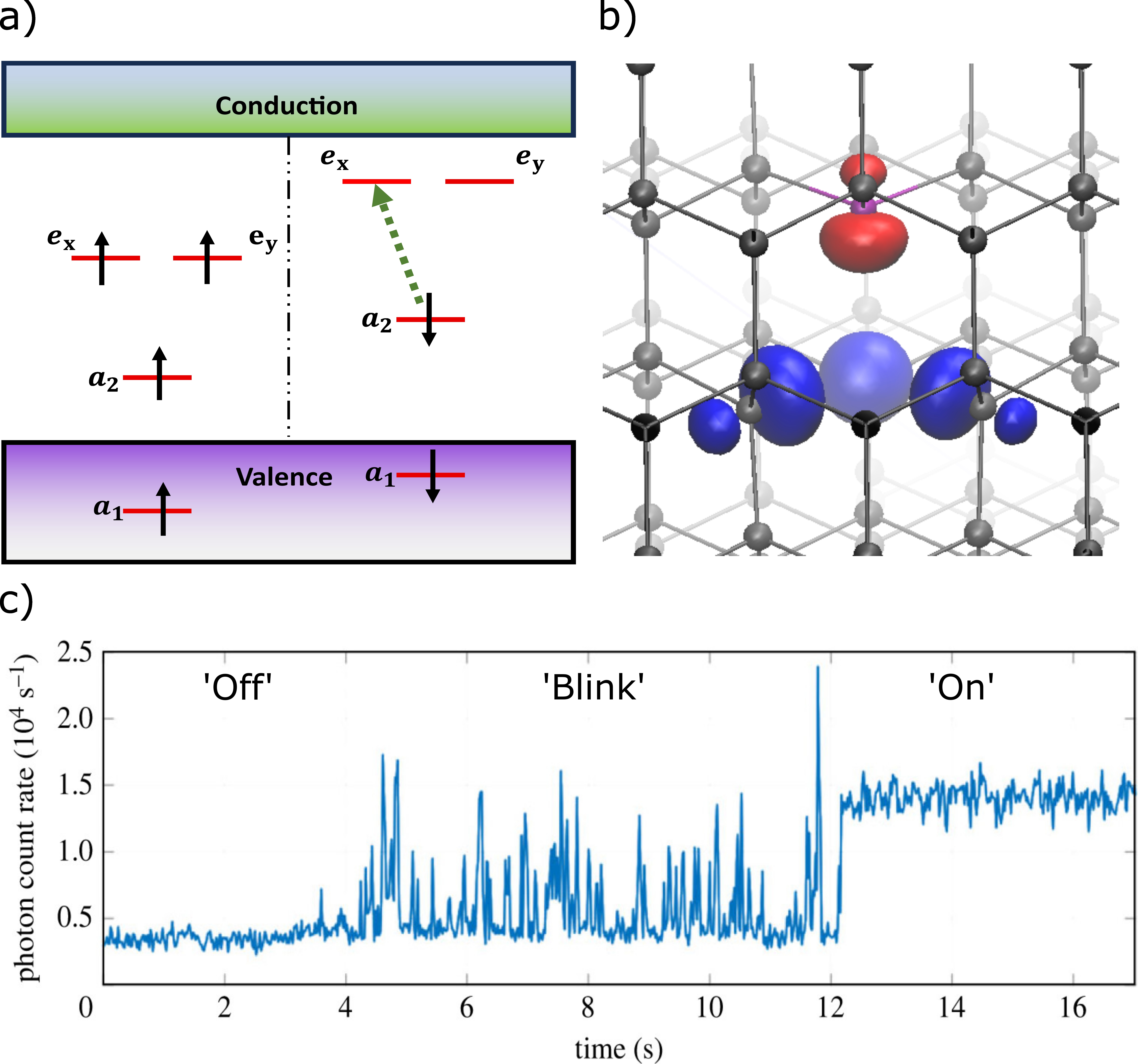}
  \caption{\textbf{An overview of the electronic structure of NV$^-$
  and its characteristic fluorescence signal during laser writing.}\\
  \textbf{a.} Electronic configuration of the negatively charged NV centre in diamond in its ground state. Optical absorption promotes the spin-down electron in the $a_2$ orbital to the $e_x$ or $e_y$ orbital, and fluorescence returns the electron to the $a_2$ state. \textbf{b.} Illustration of the electronic orbitals of NV$^-$ in diamond. The degenerate orbitals $e_x$ and $e_y$ are shown in blue, while the $a_2$ orbital is coloured red. \textbf{c.} Example fluorescence intensity trace recorded during a laser annealing process, showing intermittent fluorescence preceding the formation of NV$^-$. \cite{kirkpatrick2024ab}}
  \label{fig1}
\end{figure}

Figure \ref{fig1} \textbf{a.} and \textbf{b.} depict the electronic configuration of the NV$^-$ centre and the corresponding electronic orbitals respectively. Fluorescence from the isolated NV$^-$ results from the electric dipole-allowed transition from the excited state ($a_1a_2e_xe_y\bar{a}_1\bar{e}_x$ or $a_1a_2e_xe_y\bar{a}_1\bar{e}_y$) to the ground state ($a_1a_2e_xe_y\bar{a}_1\bar{a}_2$).
Figure \ref{fig1}\textbf{c.} presents an example of intermittent fluorescence observed during the formation of an NV$^-$ defect \cite{chen2019laser}. The typical formation process involves transitions between 'Off', 'Blink', and 'On' states, as indicated by the photoluminescence (PL) signal.
Since the vacancy has combined with the nitrogen to form the NV centre, the remaining C$_i$ is likely still located near the formed NV centre, affecting the stability of PL signal. In the context of interstitial carbon in diamond, the split-\(\langle 100 \rangle\) \( C_i \) configuration with neutral charge is considered one of the most stable arrangements, where two carbon atoms share the same lattice site along the [100] crystallographic direction. Rather than forming the typical tetrahedral sp\(^3\) bonds, these two carbon atoms may instead adopt a bonding arrangement that includes a $\pi$ bond, resulting from the overlap of unhybridized p orbitals.\cite{weigel1973carbon}

Previous ab initio density functional theory (DFT) modelling has shown that in the electronic ground state, the orbitals of NV$^-$ and the split-$\langle$100$\rangle$ C$_i$ can hybridise for some configurations, suggesting that the NV$^-$ fluorescence might be quenched and that the observed fluorescence intermittency could be a signature of a C$_i$ migrating in the vicinity of the colour centre \cite{kirkpatrick2024ab}. Here we extend this work in two ways, firstly by introducing an improved potential to the DFT model which will provide a more accurate prediction of hybridisation and of the C$_i$ migration dynamics, and secondly by modelling the optical transitions of $NV^--C_i$ defect complexes using time-dependent DFT (TDDFT). 

Studying interactions between defects and defect migration dynamics requires modelling of a system of order nanometres in size containing thousands of atoms, such that computationally expensive DFT and \textit{ab initio} molecular dynamics (AIMD) calculations are impractical.
To conduct this investigation effectively we therefore apply two techniques. Firstly, we utilise a machine-learning Gaussian Approximation Potential (GAP).  GAP-20 is a specific inter-atomic transferable machine learning potential designed for carbon-related materials, providing an alternative to achieve large-scale molecular dynamics simulations. Secondly, we use linear-scaling DFT. The computational cost of traditional DFT methods scales cubically with the number of atoms, whereas linear-scaling DFT offers higher computational efficiency and scalability for studying large-scale systems. 

We first introduce the construction of the GAP-CN (Carbon-Nitrogen) potential based on the existing GAP-20 potential. Then, the constructed GAP-CN potential is applied to optimize the geometry of input structures for TDDFT calculation and run molecular dynamics simulations.
The TDDFT results begin with observing the changes in the absorption spectrum of pure NV$^-$ in diamond at 300 K, where a blue shift of around 5 nm in wavelength is found compared to the results at 0 K.
Then, to further investigate the interaction between C$_i$ and NV$^-$, random optimized samples consisting of one C$_i$ and one NV$^-$ are constructed, followed by TDDFT calculation where 'bright', 'spike' and 'dark' states are classified based on their absorption spectrum features. Finally, a molecular dynamics(MD) simulation, assisted with the formed GAP-CN, is performed to reveal the diffusion mechanics of defects under thermal excitation. 
The MD results suggest that C$_i$ may either diffuse away from NV or recombine with V, thereby leaving pure NV alone or destroying the formed NV. In the case of diffusing away, the PL signal would transition from blinking to stable, as shown in Figure \ref{fig1} \textbf{c.}

\section{Method}

\subsection{Training procedure of GAP-CN}

\begin{figure}[H]
    \centering
    \includegraphics[scale=0.6]{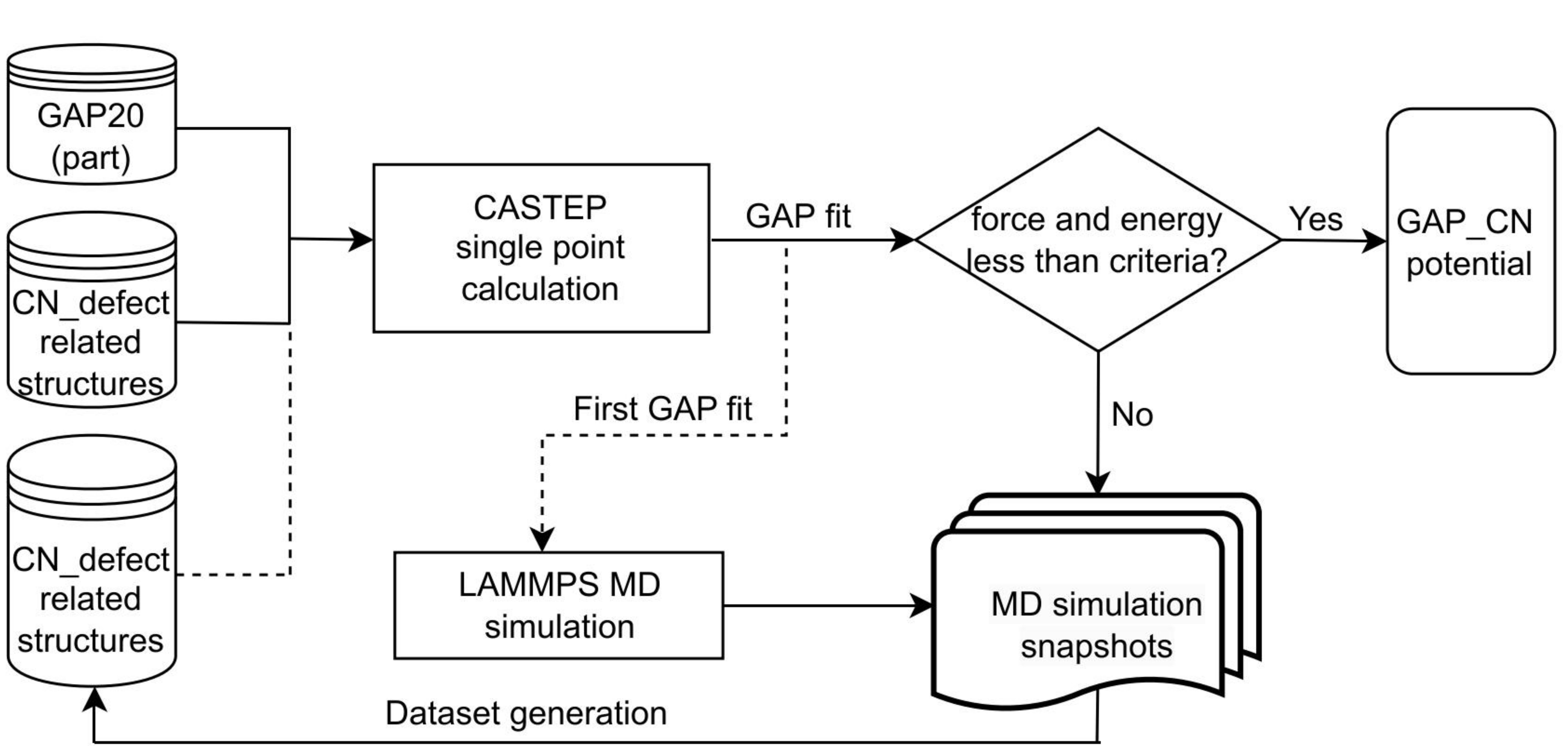}
    \caption{\textbf{A flowchart illustrating the process for generating the GAP-CN potential.}}
    \label{fig_s1}
\end{figure}
\paragraph{Training procedure of GAP-CN}
Figure \ref{fig_s1} shows the training procedure for the construction of GAP-CN. The crystalline structures from GAP20 and the CN defect-related structures serve as input geometries for DFT calculations using CASTEP.
In particular, the CN defect-related structures consist of a diamond supercell composed of $3\times3\times3$ conventional cells, containing single defects such as the nitrogen-vacancy (NV), substitutional nitrogen (N$_s$), interstitial nitrogen (NN$_i$), divacancy (VV), monovacancy (V), interstitial carbon (C$_i$), and combinations of any two of these single defects.
The total energy and atomic force data are then used to fit the parameters of the GAP-CN potential in the GAP fitting program, generating the first iteration of the GAP-CN potential. This first iteration is subsequently used for molecular dynamics simulations in LAMMPS.
To enhance the precision of force and energy calculations provided by the GAP-CN potential, frames extracted from the MD simulations are used as additional CN defect-related structures to refine the best-fit potential. The final training dataset comprises a total of 7,343 configurations.

\paragraph{Database construction}
The dataset of this potential is made up of three parts.
The first part is created based on the selected structures out of the GAP20 \cite{rowe2020accurate} dataset.
This part of the GAP-CN training dataset involves structures of 'Dimer', 'SACADA'(database of exotic carbon allotropes) \cite{hoffmann2016homo}, 'Fullerenes', 'Nanotubes', 'Crystalline Bulk', 'Defects', 'Graphite', 'Surfaces', 'Crystalline RSS'(random structure search from GAP17 potential) \cite{deringer2017extracting}, 'Graphene', 'Diamond', 'Single Atom', and 'Graphite Layer Sep'.
There are 2965 structures in total from the GAP20 database.
Next, to increase the accuracy of the ‘defective’ structures, various nitrogen-doped and carbon-related defects, and their combinations, are generated.
Here, a corresponding idea of ‘defective’ needs to be initialised in the form of a variety of defects including nitrogen-vacancy (NV), substitutional nitrogen ($N_s$), interstitial nitrogen ($N_i$), divacancy (VV), monovacancy (V) and interstitial carbon ($C_i$) and a combination of any two of them.
Specifically, we first build a $3\times3\times3$ perfect diamond supercell. Next, we randomly create the different defects within the diamond lattice. 
Finally, we perform a random perturbation on the atoms' position and the lattice parameters ranging from 10 \% to 30 \%.
The third part of our dataset is from the iteration loop to improve the accuracy of force and energy error. 

\paragraph{DFT calculation}
The synthetic samples are then applied to DFT single-point calculations based on the CASTEP code.\cite{clark2005first} The Perdew-Burke-Ernzerhof (PBE) exchange-correlation functional \cite{perdew1996rationale} is utilised with ultrasoft pseudopotentials. The cutoff energy is set to 600 eV and the Monkhorst-Pack (MP) grid spacing is specified to 0.02.

\paragraph{Training the GAP-CN potential}
After constructing the dataset through first-principles calculations, the input sets are trained into a GAP using the Quippy code.\cite{bartok2013representing}\cite{bartok2010gaussian}. The fitting program is parallelised over nodes using ScaLAPACK with MPI and OpenMP on a high-performance computing (HPC) machine\cite{klawohn2023massively}. 
Table \ref{tab6.1} summarises the GAP-CN training parameters and the three classes of descriptors used (up to 2-body, 3-body and SOAP). 
The regularisation parameters for the expected errors on energies and forces are set to 0.001 eV/atom and 0.02 eV/\AA.

\begin{table}[H]
    \centering

    \begin{tabular}{|c|c|c|c|}
        \hline
        & \textbf{2-body} 
        & \textbf{3-body} & \textbf{SOAP} \\
        \hline
        \hline
        $\delta$(eV) & 2 & 0.5 & 0.2 \\
        $r_{cut}$(\AA) & 3.7 & 3 & 3.7 \\
        $r_{\Delta}$ &  &  & 0.4 \\
        \hline
        $\sigma$(\AA) &  &  & 0.5 \\
        $n_{max}$ &  &  & 12 \\
        $l_{max}$ &  &  & 4 \\
        Sparsification & Uniform & Uniform & CUR \\
        \hline
    \end{tabular}
    \caption{Key parameters used in the CN-GAP potential training.}
    \label{tab6.1}
\end{table}

\paragraph{Molecular dynamics simulations} 
After the first trial of GAP fitting, the constructed potential is applied to perform molecular dynamics using the quip pair style implemented in LAMMPS \cite{thompson2022lammps}. In particular, the molecular dynamics simulation is conducted in the environment of a constant-particle number, constant-volume, constant-temperature (NVT) ensemble using a Nose-Hover thermostat at a temperature of 3000 K for 10 ps with the time step of 2 fs. This enables the generation of sufficient snapshots for the iteration procedure and provides adequate perturbation on the constructed defective structure. These constructed structures will serve as new 'CN\_defect related structure' dataset in Figure \ref{fig_s1}.

\subsection{TDDFT calculation}

\begin{figure}
  \centering
  \includegraphics[scale=0.77]{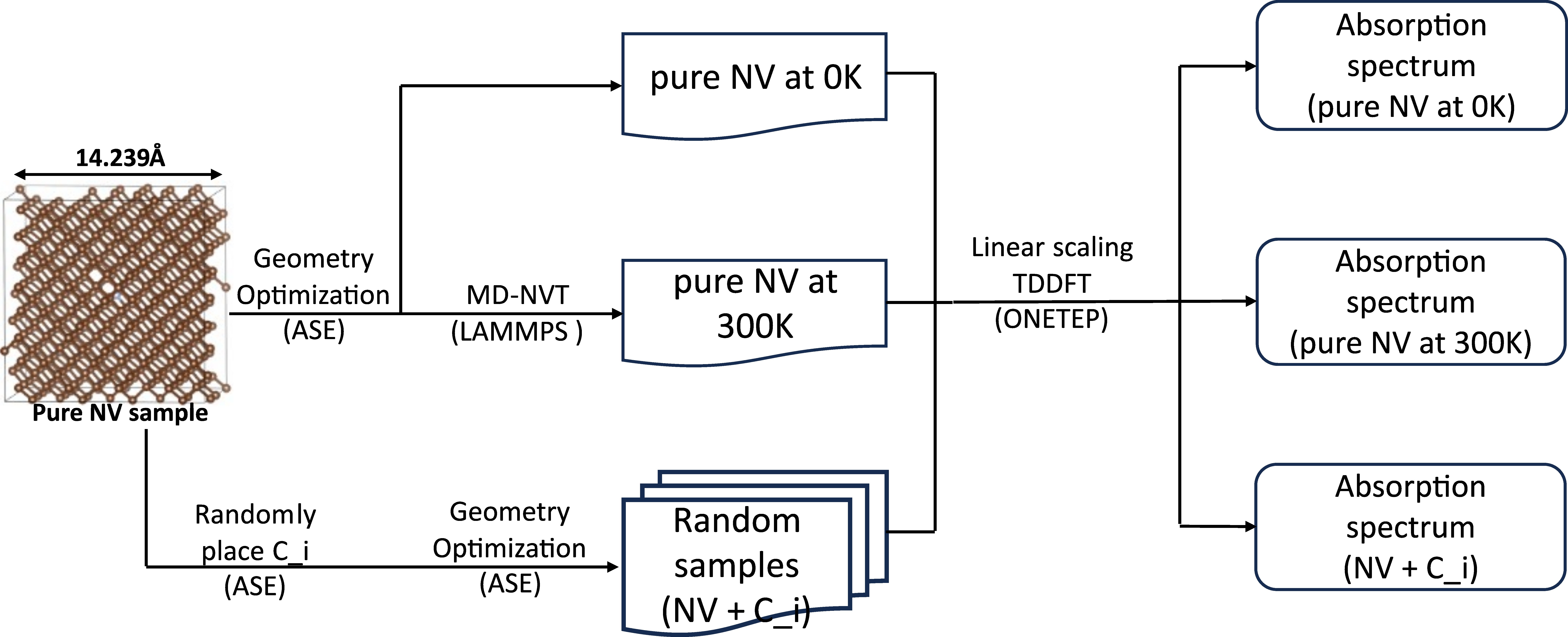}
  \caption{\textbf{Details of workflow for absorption spectrum investigation.}}
  \label{fig2}
\end{figure}
A general workflow for this section is illustrated in Figure \ref{fig2}.
The simulation begins with a pure NV structure in a supercell diamond lattice comprising 510 carbon atoms and 1 nitrogen atom. Two main tasks are then performed subsequently.
The first task focuses on investigating the differences in the excited states of NV$^-$ at 0 K and 300 K. Calculations were performed at 300 K for two primary reasons: firstly, photoluminescence (PL) measurements were conducted at room temperature; and secondly, to examine the influence of room temperature on the emission spectrum.
The 0 K structure is generated by geometry optimisation, accelerated using the GAP-CN potential via the ASE package.\cite{larsen2017atomic}
On the other hand, the 300 K pure NV$^-$ sample undergoes an additional step involving an NVT ensemble molecular dynamics simulation, from which seven snapshots are extracted at equal time intervals of 250 fs each.
All these constructed samples are then used in linear-scaling TDDFT simulations with ONETEP \cite{prentice2020onetep} to calculate absorption properties.
The second task involves comparing the excited states across various ensembles of NV$^-$ and C$_i$. The random ensemble samples are generated by placing a single carbon atom into the pure NV sample at random, followed by geometry optimisation accelerated using the GAP-CN potential. Analogously, these optimized configurations will be applied in TDDFT simulations for absorption spectrum comparison. There are 100 randomly constructed samples of NV + C$_i$ used for the TDDFT calculations.

\paragraph{Sample construction}
The input NV sample in a $4\times4\times4$ supercell diamond structure is constructed by \textit{CrystalMaker} for further process.\cite{palmer2015visualization} 
The atomic simulation environment (ASE) is employed to process geometry optimization with the constructed GAP-CN potential.
During the procedure of geometry optimization, the optimizer of BFGS\cite{fletcher2000practical} is applied with the converged criterion force on all individual atoms less than 0.001 eV/Å. There are 100 randomly constructed samples containing one $NV$ and one $C_i$ for further linear scaling TDDFT calculations.

\paragraph{Linear scaling DFT}
The aforementioned structures are then applied in ONETEP\cite{prentice2020onetep} for a single point, conduction and TDDFT calculation.
In particular, PBE exchange-correlational functionals \cite{ernzerhof1999assessment} are used for all the simulations while a pseudopotential is employed in treating the interaction between core and valence electrons. To explicitly model the real electronic structure of $NV^-$, one extra charge is added into the system and the spin-polarized feature is turned on.
The overall spin of the system is fixed appropriately as 2. 
The cutoff energy and the Non-orthogonal Generalized Wannier Functions (NGWFs) are converged with the value of 1000 eV and 9 Bohr radii. To better describe the conduction band convergence, the radius of NGWFs increases to 10 Bohr radii. Twelve conduction bands are optimized and ten TDDFT states are calculated. 

\paragraph{Molecular dynamics}
The molecular dynamics simulation for $NV^-$ at 300 K is conducted using LAMMPS (Large-scale
Atomic/Molecular Massively Parallel Simulator) with the GAP-CN potential is implemented with QUIP.\cite{thompson2022lammps}
The MD simulations are executed at 300 Kelvin using a Nose-Hoover thermostat in the NVT ensemble. Atomic velocities are initialized based on a specified temperature using a random seed, and the
simulation time step is set to 1 femtosecond and the duration time for the simulation is 2000 fs.
The temperature profile is checked to stabilize at around 300K with a less than 20 K fluctuation. Seven final snapshots are extracted at equal time intervals for subsequent linear scaling TDDFT calculations.

\subsection{Molecular dynamics for NV +C$_i$}
This simulation aims to investigate the possible diffusion mechanisms of C$_i$ around NV during laser annealing.
Two models are considered in this part: the Vacancy Side Model, where C$_i$ is initially located on the vacancy side of NV, and the Nitrogen Side Model, where C$_i$ is placed on the nitrogen side of NV.
The detailed settings of the molecular dynamics simulation are the same as those used in the previous NVT ensemble (refer to the SI), except that the temperature is adjusted to 2600 K to balance computational resources and efficiency.
Although in real laser annealing experiments, the diffusion of C$_i$ might be exciton-assisted rather than thermally driven \cite{griffiths2021microscopic}, NVT ensemble molecular dynamics simulations provide a comprehensive understanding of the possible movement of C$_i$ around NV.

\section{Results and discussion}

\subsection{Validation of the GAP-CN potential}

\begin{figure}[H]
    \centering
    \includegraphics[scale=0.75]{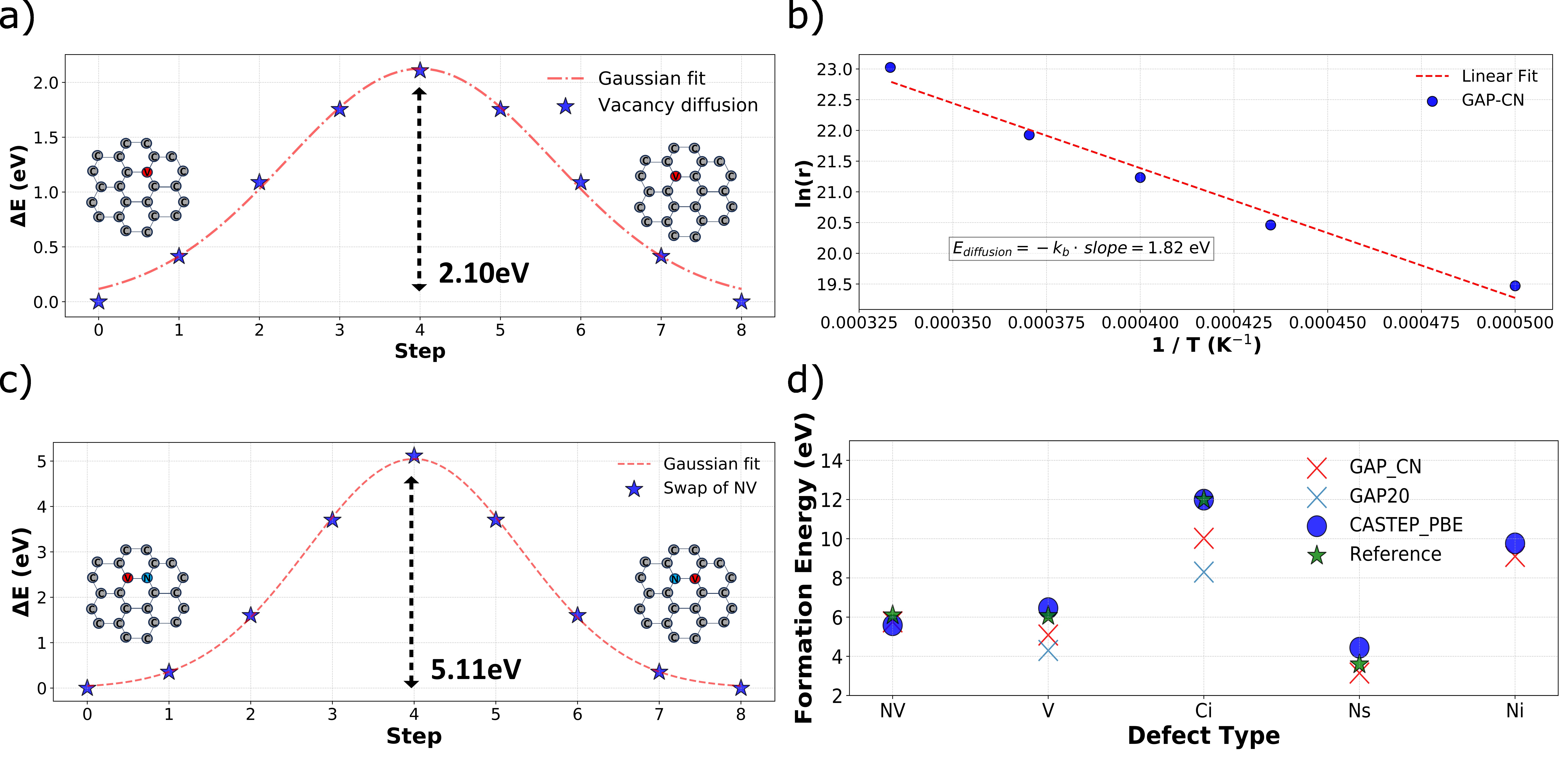}
    \caption{\textbf{A summary of the validation tests for the GAP-CN potential.}\\
    Applied the Nudged Elastic Band (NEB) method to investigate \textbf{a.} the diffusion barrier of vacancies in diamond, and \textbf{b.} the barrier for NV inversion. Graph \textbf{c.} illustrates an analysis of the diffusion barrier of interstitial carbon using a thermodynamic approach. \textbf{d.} Comparison of the formation energies computed by GAP-CN, GAP20 \cite{rowe2020accurate}, CASTEP and related reference on nitrogen-vacancy (NV) (DFT-PBE) \cite{gali2019ab}, vacancy (V) (DFT-PBE) \cite{slepetz2014divacancies}, interstitial carbon (C$_i$) (experimental results) \cite{laidlaw2020point} and substitutional nitrogen ($N_s$) (DFT-B3LYP) \cite{ferrari2018substitutional}}
    \label{fig6.4}
\end{figure}

\paragraph{Diffusion barrier comparison}
This generated potential will primarily be used to study the diffusion of defects in diamond. To this end, we recalculated the diffusion barrier energies for V, NV, and C$_i$ defects for comparison.
To investigate the vacancy
diffusion, we optimise the initial and final geometry as shown in Figure \ref{fig6.4}\textbf{a} where a single layer of diamond in the $\langle$111$\rangle$ direction is displayed. 
The diffusion barrier of a vacancy is calculated using the Python library Atomic Simulation Environment (ASE) \cite{larsen2017atomic} with the nudged elastic band (NEB) method, by tracing the potential energy of the diffusing carbon atom.
The calculated diffusion barrier is 2.1 eV (Figure \ref{fig6.4} \textbf{a}) and the experimental value is 2.3 eV \cite{davies1976optical}, which is an 8\% difference.
The inversion of the NV defect exhibits a hopping barrier of 5.11 eV across the entire system, as shown in Figure \ref{fig6.4} \textbf{b.}, with a 6\% discrepancy compared to the 4.8 eV barrier height reported in Ref. \cite{deak2014formation}.
For the investigation of the diffusion barrier of interstitial carbon, a thermal dynamics method is applied. This method is based on the Arrhenius equation (AE), expressed as follows:
\begin{equation}\label{eq:6.1}
    r = r_0 \times e^{(-\frac{E_{diffusion}}{k_{b}T})}
\end{equation}
Here, \( r \) represents the diffusion frequency, \( r_0 \) corresponds to the frequency factor, which remains constant, while \( k_b \) is the Boltzmann constant and \( T \) represents the corresponding temperature.
To initiate the simulation, the same sample containing one interstitial carbon in a \(6 \times 6 \times 6\) diamond supercell is subjected to NVT molecular dynamics simulations at temperatures of 2000, 2300, 2500, 2700, and 3000 K. The simulation employs the Nose-Hover thermostat for 10 ps with the time step of 2 fs. 
For each temperature simulation, we record the hopping frequency of interstitial carbon. Using Equation \ref{eq:6.1}, we can simplify the relationship between the diffusion frequency \( r \) and the temperature \( 1/T \) as follows:

\begin{equation}\label{eq:6.2}
    \ln(r) = -\frac{E_{diffusion}}{k_b} \times \frac{1}{T} + \ln(r_0)
\end{equation}
In this case, the diffusion barrier of interstitial carbon in diamond can be determined from the slope of the linear plot of $\ln(r)$ against \( 1/T \).
As shown in Figure \ref{fig6.4} \textbf{c.}, a linear fit is presented, and the diffusion barrier energy is calculated to be 1.8 eV. This result is consistent with the experimental findings in Ref. \cite{allers1998annealing}, which report an activation energy of 1.68 eV.
A summary of the diffusion energy comparison is provided in Table \ref{tab:diffusion_barriers}.
\begin{table}[H]
    \centering
    \begin{tabular}{|c|c|c|c|}
        \hline
        Defect & Method & Diffusion Barrier (eV) & Reference \\ 
        \hline
        NV & GAP-CN(NEB) & 5.11 & 4.8 (DFT-PBE)\cite{deak2014formation} \\ 
        C$_i$ & GAP-CN(AE) & 1.8 & 1.68 (Exp.)\cite{allers1998annealing} \\ 
        V& GAP-CN(NEB) & 2.1 & 2.3 (Exp.)\cite{davies1976optical} \\ 
        \hline
    \end{tabular}
    \caption{Comparison of Diffusion Barriers for Various Defects}
    \label{tab:diffusion_barriers}
\end{table}

\paragraph{Formation energy comparison}
This section further investigates the stability of each studied defect, including NV, V, C$_i$, N$_s$, and N$_i$. Since the entire dataset is trained in the neutral charge state, the comparison will be conducted based on neutral charge. The procedure for computing the formation energy follows equation \ref{eq6.3}, as shown below:
\begin{equation}\label{eq6.3}
    E_{f} = E_{tot} - \sum_i n_i E_i
\end{equation}
where $n_i$ is the number of atoms of element \(i\) in the system and $E_i$ is the chemical potential of an atom of an element \(i\). 
For an individual carbon atom, it is calculated by the total energy of a diamond lattice structure divided by the number of carbon atoms, whereas for a single nitrogen atom, the chemical potential is determined by half the total energy of $N_2$ in a sufficiently large box with 10 \AA\space as its length.
The calculated formation energy for each defect obtained using the GAP-CN, GAP-20 potential as well as using CASTEP\cite{clark2005first} with the PBE exchange-correlation functional is presented in Fig. \ref{fig6.4} \textbf{d.}.
As shown in the figure the nitrogen-related defects NV, N$_s$ and N$_i$ display a small formation energy error, but the absolute formation energy error is 1 eV for vacancy and is approximately 2 eV for C$_i$. It can be observed that the trained GAP-CN potential shows an improvement in the formation energy error for both $V$ and C$_i$ compared to the GAP-20 potential.

\subsection{Pure Negatively charged NV centre in diamond}

\begin{figure}[H]
  \centering
  \includegraphics[scale=0.6]{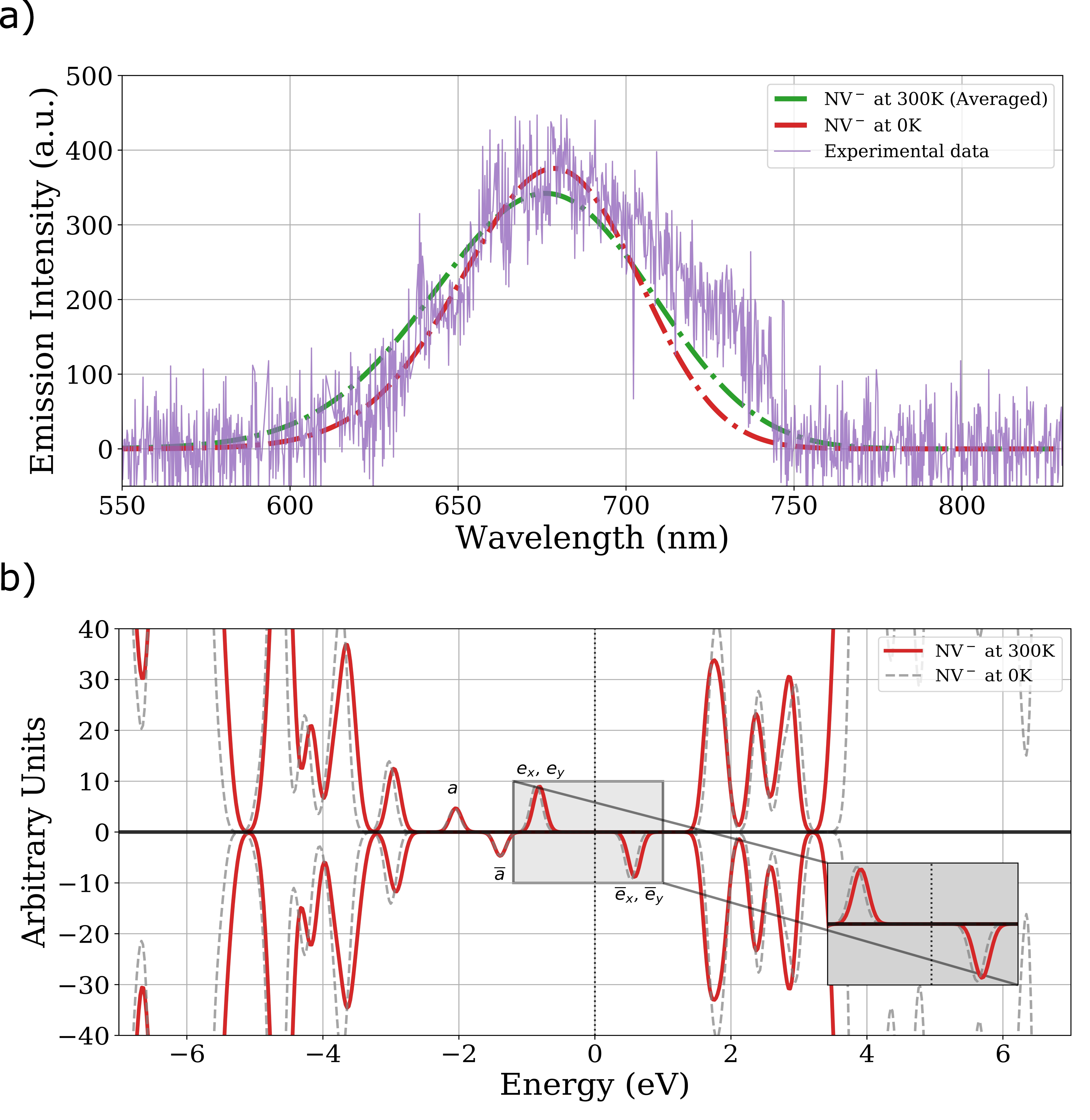}
  \caption{
  \textbf{Results of pure negatively charged NV centre in diamond.}\\
  \textbf{a.} A comparison of emission spectrum on NV$^-$ centre between experimental data and TDDFT simulation at both 0 K and 300 K. Specifically, the simulated NV$^-$ centre at 0 K is marked in red, while the averaged spectrum at 300 K is coloured green. \textbf{b.} A snapshot at 300 K is extracted to compare the density of states of NV$^-$ at 0 K.}
  \label{fig3}
\end{figure}

In this section, the NV defect is optimized using the constructed GAP-CN potential, which serves as the input for subsequent TDDFT calculations. The training dataset for GAP-CN consists entirely of neutral configurations, while the TDDFT calculations include one negative charge. Notably, the addition of this extra charge does not significantly affect the atomic configuration of the structure. A comparison of positional variations for NV optimized by GAP-CN, NV$^0$ optimized by CASTEP-PBE, and NV$^-$ optimized by CASTEP-PBE is provided in the SI, showing that the largest variation is less than 0.02~\AA.
We can therefore assume that the configurations sampled from the neutral system are a good approximation to those that would be sampled from the charged system, justifying our use of the neutral system in our MD and geometry optimizations.

The results of pure NV$^-$ at 300K demonstrate a tiny split on the \textit{$e_x$, $e_y$} orbitals which can be found on one example snapshot of its corresponding DOS graph in Figure \ref{fig3}\textbf{b.}. Specifically, the energy splitting of the two degenerated orbitals is around 0.08 eV. This implies that the thermal effect on lattice distortion at 300K would potentially split the excited states on \textit{$e_x$, $e_y$} orbitals, thus affecting their transition properties. 
According to the Franck-Condon theory, if the potential energy surfaces of the ground and excited states are approximately harmonic, the absorption and emission spectra will be symmetric about the ZPL.
In this case, the simulated absorption spectrum of pure NV$^-$ at both 0K and 300K can be transformed into the emission spectrum by applying a mirror symmetry to the zero-phonon line (ZPL) located at a wavelength of 637 nm. The comparison between experimental data and simulation can be found in graph \ref{fig3}\textbf{\textbf{a}}. 
It can be observed from the graph that the emission spectrum of the pure \( NV^- \) at both 0 K and 300 K aligns well with the experimental data at the peak of the emission spectrum, validating the simulation results.
Meanwhile, the tiny splitting of the degenerate orbitals would result in a blue shift in the emission spectrum with a value of around 5 nm. This implies that the thermal influence at 300~K is not strong enough to affect the PL signal since its emission can still be detected.

\subsection{Negatively charged NV centre and interstitial carbon}

In this section, the optimized NV + C$_i$ structure is used in TDDFT calculations to investigate the blinking phenomenon observed in the PL experimental data. The simulation results are categorized as 'Bright,' 'Spike,' and 'Dark' based on their absorption spectra, compared to the absorption spectrum of pure NV in diamond obtained from TDDFT simulations.
It is worth noting that TDDFT with the PBE functional typically underestimates the transition energy for the ground-state transition from $^3A_2$ to $^3E$ by approximately 0.1~eV, as reported in Ref.~\cite{jin2022vibrationally}. This discrepancy results in a slight blue shift in the absorption spectrum, corresponding to a wavelength difference of approximately 20--30 nm. Such an adjustment aligns better with the absorption process, as indicated by our results, where the pump laser, operating at 532~nm, appears to correspond to a lower wavelength range.

\begin{figure}[H]
  \centering
  \includegraphics[scale=0.77]{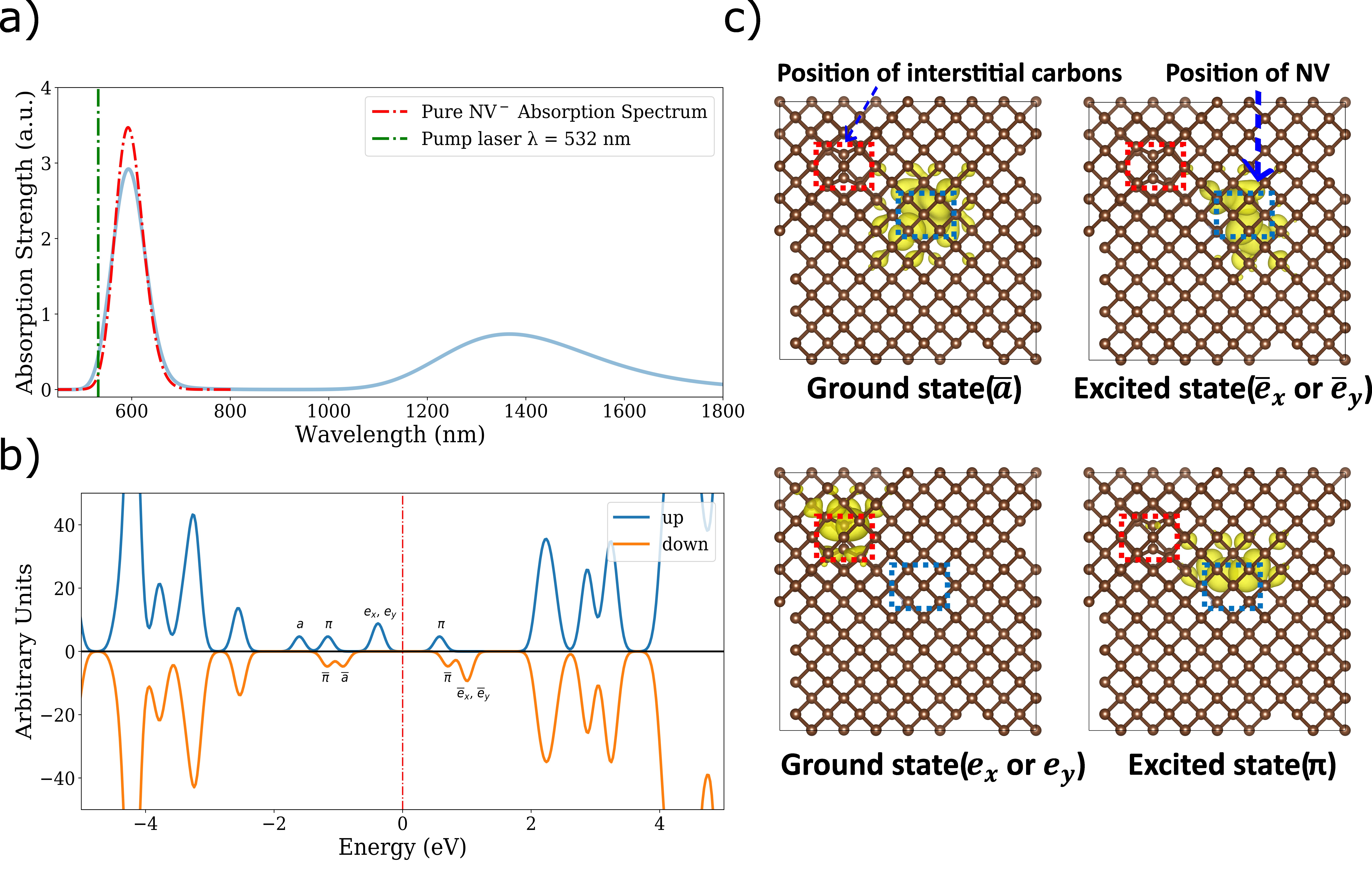}
  \caption{
  \textbf{An example of results for non-hybridization (Bright).}\\
  \textbf{a.} An illustration of the absorption spectrum on all the non-hybridized situations. \textbf{b.} A featured density of state (DOS) plot on the non-hybridized situation. \textbf{c.} A demonstration of transition on the electronic orbitals of the ground state ($\bar{a}$) to the excited states ($\bar{e}_x$ or $\bar{e}_y$) of the NV centre at the wavelength of around 600~nm and a transition from the ground state (${e}_x$ or ${e}_y$) to the excited states ($\pi$) of interstitial carbon whose wavelength is around 1400~nm.}
  \label{fig4}
\end{figure}

\paragraph{Non-hybridization(Bright)} Non-hybridization means the orbitals of NV$^-$ and interstitial carbon can be analyzed separately. In our previous study \cite{kirkpatrick2024ab}, it was shown that non-hybridization implies no effect on NV emission, whereas hybridization is likely to influence NV emission. An illustration of the absorption spectrum for an example of the non-hybridized situations can be found in Figure \ref{fig4} \textbf{a.} where two main peaks can be addressed at around 1370~nm and 600~nm. To further investigate the transition behind these peaks, their corresponding hole states and electron states are present in Figure \ref{fig4} \textbf{c.}. Meanwhile, a typical DOS from the non-hybridized situation is presented in graph \ref{fig4} \textbf{b.}. Combining the DOS plot and hole-electron orbitals, the absorption peak located at around 1370~nm refers to the transition from NV$^-$'s \textit{$e_x$, $e_y$} manifolds to C$_i$'s \textit{$\pi$} bond in the spin-up configuration while the peak at 600~nm can be interpreted as a localised transition on the  NV$^-$ colour centre which is  \textit{a} orbital to \textit{$e_x$, $e_y$} orbitals.
Since the localised transition from NV$^-$ remains and there is no obvious shift on its corresponding absorption peak, it can be concluded that in non-hybridization, the absorption procedure during experiments will not be affected significantly.

\begin{figure}[H]
  \centering
  \includegraphics[scale=0.7]{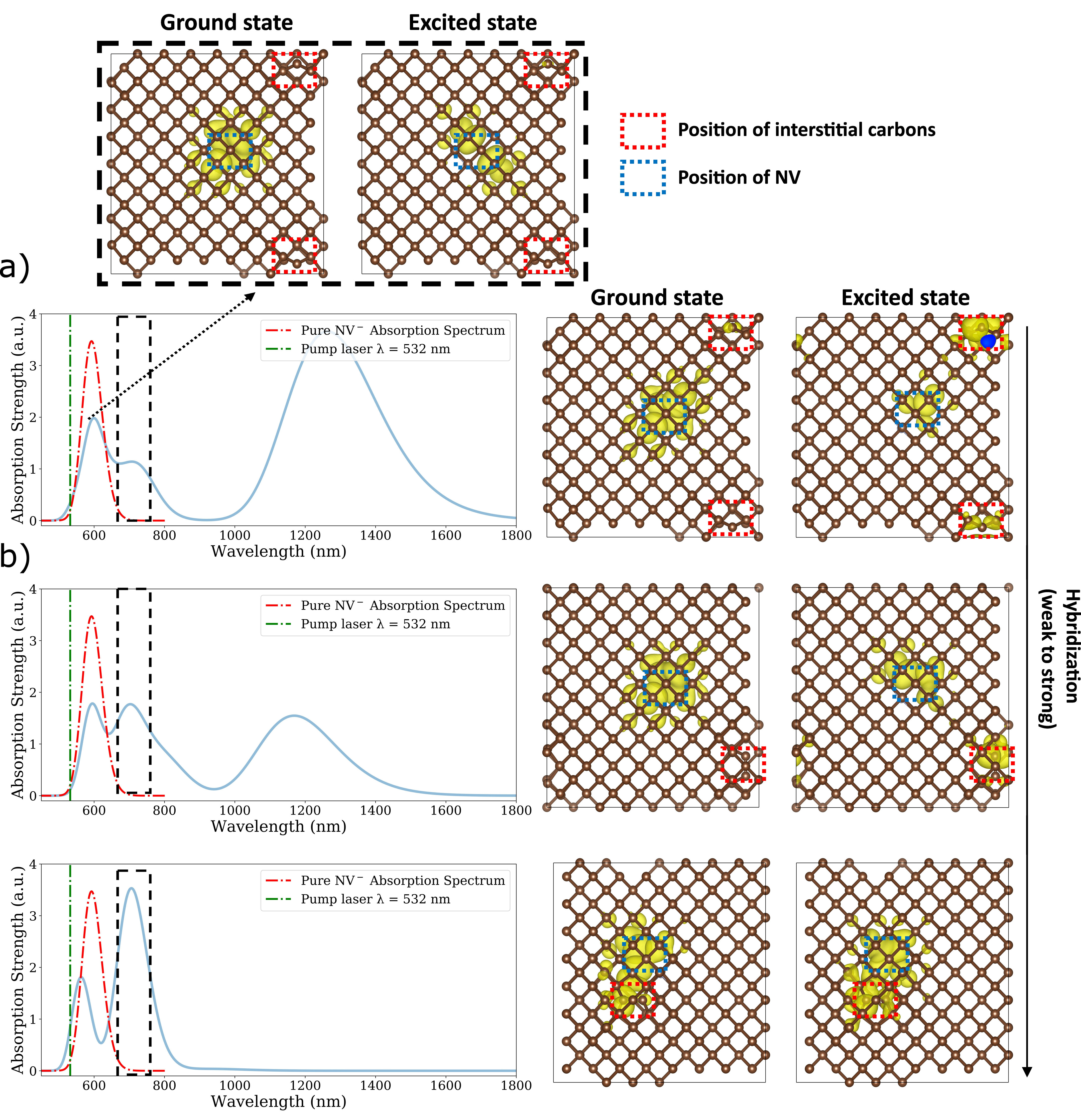}
  \caption{
  \textbf{An example of results for hybridization (Spike).}\\
  \textbf{a.} An example of weak-hybridized on both the absorption spectrum and its corresponding transition manifolds. \textbf{b.} An illustration of progression in absorption spectrum from weak to strong hybridized situation and its corresponding hole-electron states}
  \label{fig5}
\end{figure}

\paragraph{Hybridization(Spike)} In this situation, a hybridized state between NV$^-$ centre and C$_i$ transition is detected at the wavelength around 700~nm and the transition from NV$^-$ will be weaker as the increase of hybridized component between NV$^-$ centre and C$_i$. This tendency can be observed from Figure \ref{fig5} where the absorption peak from around 600~nm decreases gradually as the absorption strength at 700~nm becomes stronger. Details of the absorption peak interpretations can be found in Figure \ref{fig5} \textbf{a.} where an example of weak-hybridization is present. 
In this graph, there are three main absorption peaks at around 600~nm, 700~nm and 1300~nm. Because the excited pump laser is at the wavelength of 532 nm, marked as a dashed line in green, the absorption peak localized at 1300 nm may not be transferred to the excited state. 
The transition at 600~nm can be regarded as the transition within NV$^-$ centre from \textit{$\bar{a}$} to \textit{$\bar{e}_x$, $\bar{e}_y$} orbitals despite the weakening of absorption strength by comparing to the pure original NV$^-$ absorption peak. 
An interesting effect happens at the peak from 700 nm, which corresponds to a transition from the  \textit{$\bar{a}$} state of the NV$^-$ to a 'weak-hybridized' state which is a hybridisation of the NV$^-$ centre's \textit{$\bar{e}_x$, $\bar{e}_y$} states and the C$_i$'s $\pi$ state.
This 'weak-hybridized' phenomenon could directly reduce the transition intensity at 600~nm, potentially accounting for the blinking signal observed during the fluorescence procedure.
Furthermore, with the increase of hybridization, the transition at 700~nm will be stronger whereas the transition at approximately 600~nm will weaken and vanish eventually. 
This suggests that the absorption will weaken and potentially disappear altogether in the case of strong hybridization.

To better understand the progress of hybridization and its influence on the transition within NV$^-$ centre, further absorption spectrum analysis is performed on three different levels of hybridization, shown in Figure \ref{fig5}. As can be extracted from the graph, with the increase of absorption strength for the 700~nm peak, defined as the signal of hybridization, the 600~nm transition localised on the NV$^-$ centre weakens gradually. 
The corresponding electron-hole configuration for those marked absorption peaks is depicted on the right, where the volume of the isosurface around interstitial carbon highly determines the level of hybridization intensity. 
Another key finding from the graph is that for the bottom situation, all the orbitals from interstitial carbon are fully hybridized with NV$^-$ centre, presenting no peaks around 1300~nm.

\begin{figure}[H]
  \centering
  \includegraphics[scale=0.75]{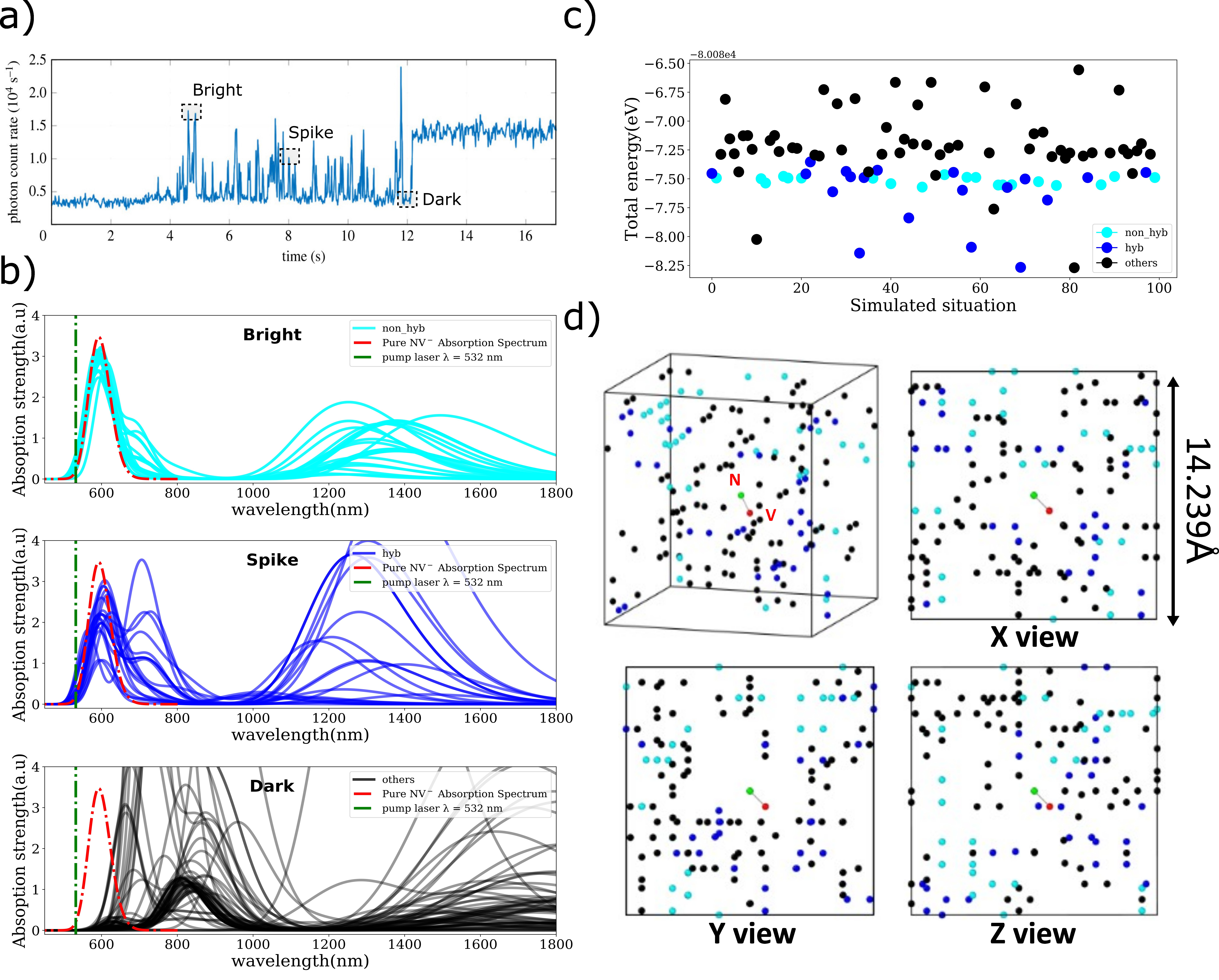}
  \caption{\textbf{A summary of all the simulated cases on one negatively charged NV and one interstitial carbon.}\\
  \textbf{a.} An illustration of PL signal in the experiment.\textbf{b.} An illustration of 'Bright', 'Spike' and 'Dark' states together with their corresponding absorption spectrum.  \textbf{c.} Total energy variance for the simulated situations. \textbf{d.} Spatial distribution for the classified situations. The NV$^-$ centre is located at the centre of the supercell, while the rest corresponds to interstitial carbon where 'Bright', 'Spike' and 'Dark' locations are marked in cyan, blue and black.}
  \label{fig6}
\end{figure}

\paragraph{Others(Dark)} Except for the Non-hybridization and Hybridization scenarios, the absorption spectra of the remaining situations, defined as Others, are relatively complicated. A summary can be found in the bottom graph Figure \ref{fig6} \textbf{b.}.
Although it is not possible to systematically analyse these spectra in terms of the NV$^-$ and C$_i$ defect states, it can be observed from the graph that all of the peaks in these spectra are shifting further away from the wavelength of the pump laser at 532~nm. This indicates that there will be no excitation among all of these other situations.     
Examination of individual situations showed that this significant shift in the absorption spectrum can be caused by hybridization, energy level shifting, charge relocation, and other effects. 

\paragraph{Summary and spatial distribution analysis} After investigating some individual situations on non-hybridization, hybridization and others, the overview of the absorption spectra of all the random samples can be re-plotted by the classification, shown in Figure\ref{fig6} \textbf{b.}. In the graph, non-hybridization is marked in cyan, presenting 'bright' states in the absorption procedure while hybridization is coloured in blue, showing 'blinking' states. Finally, due to the right shift of all the 'others' situations, they are tagged in night blue, showing 'dark' states during the excitation procedure.  

The graph \textbf{c.} in Figure \ref{fig6} demonstrates the spatial distribution on C$_i$ and NV$^-$ of all the situations contained in this research. The carbons located at the pristine diamond lattice are eliminated by the Wigner-Seitz approach \cite{zou1994topological}. From our previous study \cite{kirkpatrick2024ab}, we learn that under the annealing procedure, the interstitial carbon is mobile with a possibility to diffuse around the formed NV$^-$ centre.
The total energy for each simulated configuration is displayed in Figure \ref{fig6} \textbf{b.}, revealing that the maximum energy difference between random samples is less than 2~eV while most of the variance is less than 0.5~eV. 
This implies that interstitial carbon can migrate to all of these positions, hopping between the states of 'bright', 'spike', and 'dark' during the laser annealing procedure. 

Among all the simulated situations, around 20\% of them are shown as 'bright', 20\% of them are marked as 'spike' and the rest of them are present in the 'dark' state. 
The outcomes associated with these states are significantly influenced by atomic configurations, the spatial proximity of defects and charge states.  

\subsection{Molecular dynamics simulations} 
The previous section demonstrated that when C$_i$ is localized around the NV centre, the PL signal may exhibit blinking due to the presence of hybridized orbitals between NV and C$_i$. This section presents two MD simulations using the NVT ensemble at a temperature of 2600~K, with C$_i$ positioned either on the vacancy side or the nitrogen side of NV. These simulations aim to provide a qualitative understanding of the diffusion mechanism of C$_i$ around NV, and therefore how the configurations studied in the previous sections may emerge dynamically.

\paragraph{Vacancy side model} In this simulation, the interstitial carbon is positioned at the vacancy side of the NV centre, approximately 6 \AA\ away. An overview of the simulation procedure can be found in Figure \ref{fig7} \textbf{a.} where snapshots captured at specific time intervals are presented. The carbon atoms in the regular diamond lattice have been removed using Wigner-Seitz defect analysis \cite{zou1994topological} for better visualization.
The distance of interstitial carbons from the vacancy position is plotted as a function of simulation time.
Combining the tracking snapshots with the corresponding distance data, we observe that initially, the interstitial carbon relaxes into a split-$\langle$100$\rangle$ interstitial carbon configuration, which is the predominant configuration. However, at approximately 3200~fs, one of the interstitial carbons combines with the vacancy position, leading to the disappearance of the NV centre. 
This observation is clearly present in the plot, where the distance between interstitial carbon 1 and the vacancy abruptly decreases to zero, indicating the merger of the interstitial carbon with the vacancy site.

\begin{figure}[H]
  \centering
  \includegraphics[scale=0.6]{fig9.jpg}
  \caption{\textbf{Results of molecular dynamics simulations for two different models.}\\
  \textbf{a.} An overview of NVT molecular dynamics simulation illustrates the recombination of C$_i$ and $V$ defects, particularly when the C$_i$ is initially located immediately next to the vacancy side of the NV centre. \textbf{b.}  A general graph illustrating the initial placement of an interstitial on the nitrogen side of the NV centre indicates that it will tend to diffuse away over time.}
  \label{fig7}
\end{figure}

\paragraph{Nitrogen side model} The same simulation is conducted with the initial condition of the C$_i$ being situated at the nitrogen side of the NV centre, as depicted in Figure \ref{fig7} \textbf{b.}. The displacement between these two defects is approximately 6 Å. From the figure, it is evident that within the first 2000~fs, the interstitial carbons diffuse away from the NV centre. However, after 2000~fs, the interstitial carbon diffuses around the boundary of the supercell with a noticeable tendency to diffuse back towards the NV centre.

In summary, the model suggests that if an interstitial is present on the vacancy side of the NV defect, the PL signal will evolve from 'Blink' to 'Off' since the interstitial finally recombines with the vacancy of the NV centre. However, if the C$_i$ is located on the nitrogen side of the NV centre, it will tend to diffuse away, leading to the PL evolving from 'Blink' to 'On'.

\section{Discussion}

In this work, we combine the linear-scaling DFT method together with machine-learning potential GAP to investigate the physical processes behind fluorescence intermittency prior to NV$^-$ formation. 
The study initialises with the construction of a GAP-CN potential based on the existing GAP-20 potential. This GAP-CN potential is then applied to assist in the input structure for TDDFT calculation and molecular dynamics simulations.
A summary of conclusions is listed below:

\begin{itemize}
    \item The results of TDDFT on pure NV$^-$ centre in diamond indicate a spectral shift of approximately 5 nm in emission at 300 K compared to 0 K. This implies that the thermal influence at room temperature is not strong enough to produce the spiking signal in PL measurement. 
\item Optimized random configurations containing one \( \mathrm{NV}^- \) centre and one interstitial carbon (\( C_i \)) were generated for TDDFT analysis. Based on their absorption spectra, these configurations were categorized into 'bright,' 'spike,' and 'dark' states, providing an explanation for the blinking behavior observed in PL signals.
\item Machine-learning-driven molecular dynamics simulations provided insights into potential diffusion mechanisms for \( C_i \) around the \( \mathrm{NV}^- \) centre under laser annealing. The simulations suggest that \( C_i \) may either recombine with the \( \mathrm{NV}^- \) centre, leading to its disappearance in PL measurements, or diffuse away, allowing for the formation of a \( \mathrm{NV}^- \) centre. This explains the observed intermittent loss and reappearance of \( \mathrm{NV}^- \) centres in PL data.
\end{itemize}
While our results provide one possible explanation for the blinking signal observed in photoluminescence (PL) measurements, there are limitations in our simulations that should be considered.
In the TDDFT calculations, the annealing procedure was not explicitly simulated. Under such conditions, different ensembles of defects could result in varying charge states, potentially leading to diverse spectral outcomes. Additionally, even when the system is pumped into an excited state, relaxation can occur via both radiative and non-radiative pathways. These competing processes may contribute to the observed blinking behaviour, suggesting that the dynamics of the excited state play a more complex role than captured in our current model.
Similarly, in the molecular dynamics simulations, we employed a heat-driven model to investigate diffusion mechanisms. However, in real experiments, the process is believed to be exciton-driven rather than solely thermally activated. This distinction may lead to differences in the dynamics and the underlying mechanisms of defect recombination and diffusion.

\section*{Data availability}
The datasets, trained potentials, and simulation results can be accessed on the Zenodo repository via \href{https://doi.org/10.5281/zenodo.13259544}{https://doi.org/10.5281/zenodo.13259544}.

\section*{Acknowledgments}
The authors would like to acknowledge the use of the University of Oxford Advanced Research Computing (ARC) facility in carrying out this work (http://dx.doi.org/10.5281/zenodo.22558). Additionally, we acknowledge the use of NQIT computing nodes and the Quantum Computing and Simulation Hub. We extend our thanks to Yuxing Zhou for valuable suggestions on ML potential training, to Xingrui Cheng for providing experimental data, and to Jacx Chan for offering meaningful insights for this work.

\section*{Contributions}
G.C: conceptualization, data curation, formal analysis, investigation, methodology, supervision, writing—original draft, writing—review and editing; 
J.C.A.P: formal analysis, supervision,  writing—review and editing.
J.M.S.: formal analysis, funding acquisition, project administration, resources, supervision, writing—original draft, writing—review and editing.

\section*{Competing 
 interests}
The authors declare no competing financial or non-financial interests.

\bibliographystyle{unsrt}  
\bibliography{references}  

\begin{thebibliography}{10}

\bibitem{degen2008scanning}
CL~Degen.
\newblock Scanning magnetic field microscope with a diamond single-spin sensor.
\newblock {\em Applied Physics Letters}, 92(24), 2008.

\bibitem{balasubramanian2008nanoscale}
Gopalakrishnan Balasubramanian, IY~Chan, Roman Kolesov, Mohannad Al-Hmoud, Julia Tisler, Chang Shin, Changdong Kim, Aleksander Wojcik, Philip~R Hemmer, Anke Krueger, et~al.
\newblock Nanoscale imaging magnetometry with diamond spins under ambient conditions.
\newblock {\em Nature}, 455(7213):648--651, 2008.

\bibitem{childress2013diamond}
Lilian Childress and Ronald Hanson.
\newblock Diamond nv centers for quantum computing and quantum networks.
\newblock {\em MRS bulletin}, 38(2):134--138, 2013.

\bibitem{ovartchaiyapong2014dynamic}
Preeti Ovartchaiyapong, Kenneth~W Lee, Bryan~A Myers, and Ania C~Bleszynski Jayich.
\newblock Dynamic strain-mediated coupling of a single diamond spin to a mechanical resonator.
\newblock {\em Nature communications}, 5(1):4429, 2014.

\bibitem{abobeih2022Nature}
M.~H. Abobeih, Y.~Wang, J.~Randall, S.~J.~H. Loenen, C.~E. Bradley, M.~Markham, D.~J. Twitchen, B.~M. Terhal, and T.~H. Taminiau.
\newblock Fault-tolerant operation of a logical qubit in a diamond quantum processor.
\newblock {\em Nature}, 606:884--889, 2022.

\bibitem{haque2017overview}
Ariful Haque and Sharaf Sumaiya.
\newblock An overview on the formation and processing of nitrogen-vacancy photonic centers in diamond by ion implantation.
\newblock {\em Journal of Manufacturing and Materials Processing}, 1(1):6, 2017.

\bibitem{schwartz2012effects}
Julian Schwartz, Shaul Aloni, D~Frank Ogletree, and Thomas Schenkel.
\newblock Effects of low-energy electron irradiation on formation of nitrogen--vacancy centers in single-crystal diamond.
\newblock {\em New Journal of Physics}, 14(4):043024, 2012.

\bibitem{chen2017laser}
Yu-Chen Chen, Patrick~S Salter, Sebastian Knauer, Laiyi Weng, Angelo~C Frangeskou, Colin~J Stephen, Shazeaa~N Ishmael, Philip~R Dolan, Sam Johnson, Ben~L Green, et~al.
\newblock Laser writing of coherent colour centres in diamond.
\newblock {\em Nature Photonics}, 11(2):77--80, 2017.

\bibitem{smith2019nanophotonics}
Jason~M. Smith, Simon~A. Meynell, Ania~C. Bleszynski-Jayich, and Jan Meijer.
\newblock Colour centre generation in diamond for quantum technologies.
\newblock {\em Nanophotonics}, 8(11):1889--1906, 2019.

\bibitem{chen2019laser}
Yu-Chen Chen, Benjamin Griffiths, Laiyi Weng, Shannon~S Nicley, Shazeaa~N Ishmael, Yashna Lekhai, Sam Johnson, Colin~J Stephen, Ben~L Green, Gavin~W Morley, et~al.
\newblock Laser writing of individual nitrogen-vacancy defects in diamond with near-unity yield.
\newblock {\em Optica}, 6(5):662--667, 2019.

\bibitem{weigel1973carbon}
C~Weigel, David Peak, JW~Corbett, GD~Watkins, and RP~Messmer.
\newblock Carbon interstitial in the diamond lattice.
\newblock {\em Physical Review B}, 8(6):2906, 1973.

\bibitem{kirkpatrick2024ab}
Andrew~R Kirkpatrick, Guangzhao Chen, Helen Witkowska, James Brixey, Ben~L Green, Martin~J Booth, Patrick~S Salter, and Jason~M Smith.
\newblock Ab initio study of defect interactions between the negatively charged nitrogen vacancy centre and the carbon self-interstitial in diamond.
\newblock {\em Philosophical Transactions of the Royal Society A}, 382(2265):20230174, 2024.

\bibitem{rowe2020accurate}
Patrick Rowe, Volker~L Deringer, Piero Gasparotto, G{\'a}bor Cs{\'a}nyi, and Angelos Michaelides.
\newblock An accurate and transferable machine learning potential for carbon.
\newblock {\em The Journal of Chemical Physics}, 153(3), 2020.

\bibitem{hoffmann2016homo}
Roald Hoffmann, Artyom~A Kabanov, Andrey~A Golov, and Davide~M Proserpio.
\newblock Homo citans and carbon allotropes: for an ethics of citation.
\newblock {\em Angewandte Chemie International Edition}, 55(37):10962--10976, 2016.

\bibitem{deringer2017extracting}
Volker~L Deringer, G{\'a}bor Cs{\'a}nyi, and Davide~M Proserpio.
\newblock Extracting crystal chemistry from amorphous carbon structures.
\newblock {\em ChemPhysChem}, 18(8):873--877, 2017.

\bibitem{clark2005first}
Stewart~J Clark, Matthew~D Segall, Chris~J Pickard, Phil~J Hasnip, Matt~IJ Probert, Keith Refson, and Mike~C Payne.
\newblock First principles methods using castep.
\newblock {\em Zeitschrift f{\"u}r kristallographie-crystalline materials}, 220(5-6):567--570, 2005.

\bibitem{perdew1996rationale}
John~P Perdew, Matthias Ernzerhof, and Kieron Burke.
\newblock Rationale for mixing exact exchange with density functional approximations.
\newblock {\em The Journal of chemical physics}, 105(22):9982--9985, 1996.

\bibitem{bartok2013representing}
Albert~P Bart{\'o}k, Risi Kondor, and G{\'a}bor Cs{\'a}nyi.
\newblock On representing chemical environments.
\newblock {\em Physical Review B—Condensed Matter and Materials Physics}, 87(18):184115, 2013.

\bibitem{bartok2010gaussian}
Albert~P Bart{\'o}k, Mike~C Payne, Risi Kondor, and G{\'a}bor Cs{\'a}nyi.
\newblock Gaussian approximation potentials: The accuracy of quantum mechanics, without the electrons.
\newblock {\em Physical review letters}, 104(13):136403, 2010.

\bibitem{klawohn2023massively}
Sascha Klawohn, James~R Kermode, and Albert~P Bart{\'o}k.
\newblock Massively parallel fitting of gaussian approximation potentials.
\newblock {\em Machine Learning: Science and Technology}, 4(1):015020, 2023.

\bibitem{thompson2022lammps}
Aidan~P Thompson, H~Metin Aktulga, Richard Berger, Dan~S Bolintineanu, W~Michael Brown, Paul~S Crozier, Pieter~J in't Veld, Axel Kohlmeyer, Stan~G Moore, Trung~Dac Nguyen, et~al.
\newblock Lammps-a flexible simulation tool for particle-based materials modeling at the atomic, meso, and continuum scales.
\newblock {\em Computer Physics Communications}, 271:108171, 2022.

\bibitem{larsen2017atomic}
Ask~Hjorth Larsen, Jens~J{\o}rgen Mortensen, Jakob Blomqvist, Ivano~E Castelli, Rune Christensen, Marcin Du{\l}ak, Jesper Friis, Michael~N Groves, Bj{\o}rk Hammer, Cory Hargus, et~al.
\newblock The atomic simulation environment—a python library for working with atoms.
\newblock {\em Journal of Physics: Condensed Matter}, 29(27):273002, 2017.

\bibitem{prentice2020onetep}
Joseph~CA Prentice, Jolyon Aarons, James~C Womack, Alice~EA Allen, Lampros Andrinopoulos, Lucian Anton, Robert~A Bell, Arihant Bhandari, Gabriel~A Bramley, Robert~J Charlton, et~al.
\newblock The onetep linear-scaling density functional theory program.
\newblock {\em The Journal of chemical physics}, 152(17), 2020.

\bibitem{palmer2015visualization}
David~C Palmer.
\newblock Visualization and analysis of crystal structures using crystalmaker software.
\newblock {\em Zeitschrift f{\"u}r Kristallographie-Crystalline Materials}, 230(9-10):559--572, 2015.

\bibitem{fletcher2000practical}
Roger Fletcher.
\newblock {\em Practical methods of optimization}.
\newblock John Wiley \& Sons, 2000.

\bibitem{ernzerhof1999assessment}
Matthias Ernzerhof and Gustavo~E Scuseria.
\newblock Assessment of the perdew--burke--ernzerhof exchange-correlation functional.
\newblock {\em The Journal of chemical physics}, 110(11):5029--5036, 1999.

\bibitem{griffiths2021microscopic}
Benjamin Griffiths, Andrew Kirkpatrick, Shannon~S Nicley, Rajesh~L Patel, Joanna~M Zajac, Gavin~W Morley, Martin~J Booth, Patrick~S Salter, and Jason~M Smith.
\newblock Microscopic processes during ultrafast laser generation of frenkel defects in diamond.
\newblock {\em Physical Review B}, 104(17):174303, 2021.

\bibitem{davies1976optical}
Gordon Davies and MF~Hamer.
\newblock Optical studies of the 1.945 ev vibronic band in diamond.
\newblock {\em Proceedings of the Royal Society of London. A. Mathematical and Physical Sciences}, 348(1653):285--298, 1976.

\bibitem{deak2014formation}
Peter De{\'a}k, B{\'a}lint Aradi, Moloud Kaviani, Thomas Frauenheim, and Adam Gali.
\newblock Formation of nv centers in diamond: A theoretical study based on calculated transitions and migration of nitrogen and vacancy related defects.
\newblock {\em Physical review B}, 89(7):075203, 2014.

\bibitem{allers1998annealing}
Lars Allers, Alan~T Collins, and Jonathan Hiscock.
\newblock The annealing of interstitial-related optical centres in type ii natural and cvd diamond.
\newblock {\em Diamond and related materials}, 7(2-5):228--232, 1998.

\bibitem{gali2019ab}
{\'A}d{\'a}m Gali.
\newblock Ab initio theory of the nitrogen-vacancy center in diamond.
\newblock {\em Nanophotonics}, 8(11):1907--1943, 2019.

\bibitem{slepetz2014divacancies}
Brad Slepetz and Miklos Kertesz.
\newblock Divacancies in diamond: a stepwise formation mechanism.
\newblock {\em Physical Chemistry Chemical Physics}, 16(4):1515--1521, 2014.

\bibitem{laidlaw2020point}
FHJ Laidlaw, Richard Beanland, David Fisher, and Phil~L Diggle.
\newblock Point defects and interstitial climb of 90° partial dislocations in brown type iia natural diamond.
\newblock {\em Acta Materialia}, 201:494--503, 2020.

\bibitem{ferrari2018substitutional}
Anna~Maria Ferrari, Simone Salustro, Francesco~Silvio Gentile, William~C Mackrodt, and Roberto Dovesi.
\newblock Substitutional nitrogen in diamond: A quantum mechanical investigation of the electronic and spectroscopic properties.
\newblock {\em Carbon}, 134:354--365, 2018.

\bibitem{jin2022vibrationally}
Yu~Jin, Marco Govoni, and Giulia Galli.
\newblock Vibrationally resolved optical excitations of the nitrogen-vacancy center in diamond.
\newblock {\em npj Computational Materials}, 8(1):238, 2022.

\bibitem{zou1994topological}
PF~Zou and RFW Bader.
\newblock A topological definition of a wigner--seitz cell and the atomic scattering factor.
\newblock {\em Acta Crystallographica Section A: Foundations of Crystallography}, 50(6):714--725, 1994.

\end{thebibliography}
\newpage
\appendix

\end{document}